\documentclass[amssymb,prd,aps,showpacs,nofootinbib]{revtex4}
\usepackage{graphicx}
\usepackage{amsmath}%
\usepackage{amsfonts}
\newcommand{\be}{\begin{equation}}
\newcommand{\ee}{\end{equation}}
\newcommand{\bq}{\begin{eqnarray}}
\newcommand{\eq}{\end{eqnarray}}
\begin{document}

\title{\textbf{Noncommutative quantum mechanics and Bohm's ontological
interpretation}}
\author{G. D. Barbosa}
\email{gbarbosa@cbpf.br}
\author{N. Pinto-Neto}
\email{nelsonpn@cbpf.br}
\affiliation{Centro Brasileiro de Pesquisas F\'{\i}sicas, CBPF, Rua Dr.
Xavier Sigaud 150 , 22290-180, Rio de Janeiro, Brazil }
\begin{abstract}
We carry out an investigation into the possibility of developing a Bohmian
interpretation based on the continuous motion of point particles for
noncommutative quantum mechanics. The conditions for such an interpretation to
be consistent are determined, and the implications of its adoption for
noncommutativity are discussed. A Bohmian analysis of the noncommutative
harmonic oscillator is carried\ out in detail. By studying the particle motion
in the oscillator orbits, we show that small-scale physics can have influence
at large scales, something similar to the IR-UV mixing.
\end{abstract}
\maketitle

\section{Introduction}

The natural appearance of noncommutativity of the canonical type in string
theory \cite{1}\ has been motivating an intensive investigation of its
implications for quantum field theory and quantum mechanics \cite{2,3}. The
theoretical\ relevance of this new and growing branch of physics was soon
recognized, since it\ gives us the opportunity to understand very interesting
phenomena. Among them are nonlocality and IR-UV mixing \cite{5}, new physics
at very short distances \cite{3,6}, and the possible implications of Lorentz
violation \cite{7}. From the experimental point of view, a great deal
of\ effort has been devoted to the search for evidence of possible
manifestations of noncommutative effects in cosmology and high-energy and
low-energy experiments\ \cite{8}. Noncommutative quantum mechanics (NCQM) has
also been an issue of great interest (see, e.g., \cite{8.5}-\cite{9}). In
addition to its possible phenomenological relevance \cite{8.7}, the study of
NCQM is motivated by the opportunity it gives us\ to understand problems that
are present in noncommutative quantum field theory (NCQFT), and perhaps in
string theory, in a framework easy to handle \cite{8.8}-\cite{9}.

In previous work \cite{9}, a new interpretation for the canonical commutation
relation consideration,%
\begin{equation}
\lbrack\widehat{X}^{\mu},\widehat{X}^{\nu}]=i\theta^{\mu\nu},\label{1}%
\end{equation}
was proposed. According to the point of view exposed there, it is possible to
interpret the commutation relation (\ref{1}) as a property of the particle
coordinate observables, rather than of the\ spacetime coordinates. This fact
was shown to have implications for the way of performing the calculations of
NCQFT and enforced a reinterpretation of the meaning of the wave function in NCQM.

The aim of this work is to investigate the possibility of developing a Bohmian
interpretation \cite{9.3} for NCQM. We shall benefit from the ideas presented
in \cite{9} to develop a deterministic theory of hidden variables that exhibit
canonical noncommutativity (\ref{1}) between the particle position
observables. Presently, there are several motivations for the reconsideration
of hidden-variable theories (see, e.g., \cite{9.4}). We are now sure that the
Copenhagen interpretation is not the unique framework where quantum phenomena
can be described. Many theoreticians consider it more as a provisory set of
rules than as the fundamental theory of quantum physics. Alternative points of
view have been proposed that claim to solve some alleged difficulties of the
Copenhagen interpretation \cite{17}.

It is a result of Bell \cite{17,16.5} that any hidden-variables model that
leads to the same results predicted by quantum theory must itself be
nonlocal.\footnote{Recently it was argued by t'Hooft \cite{9.4} that this
result is not valid at the Planck scale. The theories under consideration in
this work do not include such a possibility, since the scales considered here
are larger than the Plankian one.} Historically, the possibility that a
measurement process at one point can have an immediate effect at a point
separated in space was uncomfortable for physicists, and totally abhorrent to
Einstein in particular \cite{16.5}. For long time, local models were by far
the preferred ones, but nowadays there is no fundamental reason for that.
Noncommutative theories, as well as string theory, have been shown not to be
local, at least in the usual sense \cite{5,17.5}. Thus, as long as one
considers a noncommutative theory, the (historical) objection against
hidden-variables models for their nonlocal character is clearly senseless. In
reality, the study of noncommutative theories using hidden variables is
strongly motivated by the detailed information such a description can\ provide
us if compared to the one of the\ Copenhagen interpretation. Currently,
however, there is a lack of investigation in this direction, and we are aware
of just one work combining noncommutative geometry with a
hidden-variables\ model \cite{9.5}, where stochastic quantization is employed.

Among the hidden-variables theories, the Bohmian one occupies a distinguished
position. It has been an object of intensive investigation and application in
a wide range of branches of physics,\ like quantum field theory \cite{11}, the
phenomenology of high-energy physics \cite{12}, condensed matter and
atomic-molecular physics \cite{14}, among others \cite{15}. The enormous
resurgence of interest in the Bohmian interpretation comes from multiple
directions. From the experimental point of view, the possibility is under
consideration of testing the limits and discussing the foundations of quantum
theory in the realm\ of condensed matter and atomic-molecular\ physics
\cite{16,16.5}. Techniques using ion traps may allow information on the
behavior of individual particles to be obtained \cite{16.5}. This makes the
Bohmian formulation using particle trajectories especially attractive for
investigation, since it allows the description of individual systems.
Moreover,\ the applicability of this interpretation transcends the discussion
of the foundations of quantum theory, since the Bohmian formalism can be
adopted for ``non-Bohmian'' physicists as a tool to get intuition about the
nature of quantum phenomena by the detailed description of the underlying
dynamics it provides (see, e.g., \cite{14}). In the theoretical framework,
there is a large number of phenomena that do not fit comfortably within the
standard operator quantum formalism of the orthodox Copenhagen interpretation.
Among them, we quote dwell and tunneling times \cite{18}, escape times and
escape positions \cite{19}, and scattering theory \cite{20}. They are easily
handled by Bohm's ontological approach \cite{20.5}.

The main motivation to develop a\ Bohmian interpretation for NCQM in this work
is the variety of evidence indicating that noncommutativity must, in some way,
be related to a quantum theory of gravitation (see, e.g., \cite{2,3,4,22.5});
consequently, it may have implications for quantum cosmology \cite{20.7}. The
inadequacy of the application of the Copenhagen interpretation for quantum
cosmology has been stressed for a long time by many prominent physicists, like
Feynman\ \cite{23} (a review of the subject may be found in \cite{23.2}). As
an alternative to the Copenhagen interpretation, the Bohmian one is employed
in several works of quantum cosmology (see \cite{23.5} and references
therein). Thus, it is important to investigate, having in mind future
applications for this area, if canonical noncommutativity is compatible with
the Bohmian interpretation of quantum theory.

The organization of this work is the following. In Sec. II, we summarize the
essential concepts of NCQM\ and develop Bohmian noncommutative quantum
mechanics (BNCQM). After an informal presentation of the construction of the
theory of motion, we formalize it in a simple and compact form. An application
of the theory for the noncommutative harmonic oscillator is presented in Sec.
III. Finally, in Sec.IV, we end up with a general discussion and\ a summary of
the main results.

\section{Bohmian Interpretation for NCQM}

\subsection{Background on NCQM}

The essential NCQM necessary for this work (for details see \cite{9})\ is
summarized in what follows. According to the Weyl quantization procedure
\cite{2,3}, the realization of the commutation relation (\ref{1}) between
position observables is given by the Moyal star product defined as below:%

\begin{align}
\left(  f\star g\right)  (x) &  =\frac{1}{(2\pi)^{n}}%
{\displaystyle\int}
d^{n}kd^{n}pe^{i(k_{\mu}+p_{\mu})x^{\mu}-\frac{i}{2}k_{\mu}\theta^{\mu\nu
}p_{\nu}}f(k)g(p)\nonumber\\
&  =e^{\left(  \frac{i}{2}\theta^{\mu\nu}\frac{\partial}{\partial\xi^{\mu}%
}\frac{\partial}{\partial\eta^{\nu}}\right)  }f(x+\xi)g(x+\eta)\mid_{\xi
=\eta=0}.\label{2}%
\end{align}
The commutative coordinates $x^{i}$ are called the Weyl symbols of position
operators ${\hat{X}}^{i}$, and, if the interpretation for canonical
noncommutativity of \cite{9} is adopted, they can be considered as spacetime
coordinates. In this work we shall assume that $\theta^{oi}=0$. The Hilbert
space of states of NCQM\ can consistently be taken as the same as in the
commutative quantum mechanics, and the noncommutative
Schr\"{o}dinger\ equation is given by
\begin{align}
i\hbar\frac{\partial\Psi(x^{i},t)}{\partial t} &  =-\frac{\hbar^{2}}{2m}%
\nabla^{2}\Psi(x^{i},t)+V(x^{i})\star\Psi(x^{i},t)\;\nonumber\\
&  =-\frac{\hbar^{2}}{2m}\nabla^{2}\Psi(x^{i},t)+V\left(  x^{i}+i\frac
{\theta^{ij}}{2}\partial_{j}\right)  \Psi(x^{i},t).\label{14}%
\end{align}
The operators
\begin{equation}
\widehat{X}^{i}=x^{i}+\frac{i\theta^{ij}\partial_{j}}{2}\label{14.5}%
\end{equation}
are the observables that correspond to the physical positions of the
particles, and $x^{i}$ are the associated canonical coordinates.\footnote{An
intuition about the meaning of these coordinates may be found in the dipole
picture \cite{9,35}. For the case of NCQM, it would consist in considering
that, instead of a\ particle, the elementary object of the theory is a ``half
dipole'' whose extent is proportional to its canonical momentum, $\Delta
x^{i}=\theta^{ij}p_{j}/2\hbar$. One of its end points carries its mass and is
responsible for its interactions. The other extreme is empty. According to
this intuitive view, the change of variables $X^{i}=x^{i}-\theta^{ij}%
p_{j}/2\hbar$ corresponds to a change of coordinates of the interacting
extreme of the dipole $X^{i}$,\ where the corresponding physical particle is
located, to its\ empty one $x^{i}$. However, since $x^{i}$ is not an
interacting extremum, we are adopting the point particle interpretation as
preferential.} Methodologically, the NCQM formulated with Eqs. (\ref{14}) and
(\ref{14.5}) can be considered as the ``usual'' quantum mechanics with a
Hamiltonian not quadratic in momenta and ``unusual'' position operators
defined in Eq. (\ref{14.5}). From this point of view, the BNCQM developed
below can be considered as an extension of the usual Bohmian quantum mechanics
along the same lines.

Since the $\widehat{X}^{i}$ do not commute and satisfy the relation (\ref{1}),
the particles cannot be localized in a measurement process. Any attempt to
localize the particles must obey the uncertainty relation%

\begin{equation}
\Delta X^{i}\Delta X^{j}\geq\left|  \theta^{ij}\right|  /2. \label{14.6}%
\end{equation}

The expression for the definition of probability density $\rho(x^{i}%
,t)=\left|  \Psi(x^{i},t)\right|  ^{2}$ has a meaning that differs from that
of ordinary quantum mechanics. The quantity$\ \rho(x^{i},t)d^{3}x$ must be
interpreted as the probability that the system is found in a configuration
such that the canonical coordinate of the particle is contained in a volume
$d^{3}x$ around the point $\vec{x}$ at time $t$. Computation of the expected
values can be done\ in a similar way as in the usual formalism. Given
an\ arbitrary physical observable, characterized by a Hermitian operator
$\widehat{A}(\hat{x}^{i},\hat{p}^{i})$ [this naturally includes $\widehat
{A}(\widehat{X}^{i}(\hat{x}^{i},\hat{p}^{i}),\hat{p}^{i})$], its expected
value is defined as%

\begin{equation}
\langle\widehat{A}\rangle_{t}=\int d^{3}x\Psi^{\ast}\left(  x^{i},t\right)
\widehat{A}(x^{i},-i\hbar\partial_{i})\Psi\left(  x^{i},t\right)  .
\label{14.7}%
\end{equation}

A Hamilton-Jacobi formalism for NCQM\ is found by writing the wave function in
its polar form $\Psi=R\,e^{iS/\hbar}$, replacing it in Eq. (\ref{14}), and
spliting its real and imaginary parts. For the real part, we obtain%

\begin{equation}
\frac{\partial S}{\partial t}+\frac{\left(  \nabla S\right)  ^{2}}%
{2m}+V+V_{nc}+Q_{K}+Q_{I}=0. \label{15}%
\end{equation}
The three new potential terms are defined as
\begin{equation}
V_{nc}=V\left(  x^{i}-\frac{\theta^{ij}}{2\hbar}\partial_{j}S\right)
-V\left(  x^{i}\right)  , \label{16}%
\end{equation}%
\begin{equation}
Q_{K}=\operatorname{Re}\left(  -\frac{\hbar^{2}}{2m}\frac{\nabla^{2}\Psi}%
{\Psi}\right)  -\left(  \frac{\hbar^{2}}{2m}\left(  \nabla S\right)
^{2}\right)  =-\frac{\hbar^{2}}{2m}\frac{\nabla^{2}R}{R}, \label{17}%
\end{equation}
and
\begin{equation}
Q_{I}=\operatorname{Re}\left(  \frac{V\left[  x^{i}+(i\theta^{ij}%
/2)\partial_{j}\right]  \Psi}{\Psi}\right)  -V\left(  x^{i}-\frac{\theta^{ij}%
}{2\hbar}\partial_{j}S\right)  . \label{18}%
\end{equation}
\ $V_{nc}$ is the\ potential that\ accounts for the noncommutative
classical\ interactions, while $Q_{K}$ and $Q_{I}$ account\ for\ the quantum
effects. The noncommutative contributions contained in the latter two can be
split out by defining
\begin{equation}
Q_{nc}=Q_{K}+Q_{I}-Q_{c}\text{,} \label{18.3}%
\end{equation}
where
\begin{equation}
Q_{c}=-\frac{\hbar^{2}}{2m}\frac{\nabla^{2}R_{c}}{R_{c}}\text{ \ ,\ \ \ }%
R_{c}=\sqrt{\Psi_{c}^{\ast}\Psi_{c}}. \label{18.6}%
\end{equation}
$\Psi_{c}$ is the wave function obtained from the commutative Schr\"{o}dinger
equation containing the usual potential $V(x^{i})$, that is, the equation
obtained by setting $\theta^{ij}=0$ in Eq. (\ref{14}) before solving it. The
imaginary part of the Schr\"{o}dinger equation, which yields the differential
probability conservation law,
\begin{equation}
\frac{\partial R^{2}}{\partial t}+\nabla\cdot\left(  R^{2}\frac{\nabla S}%
{m}\right)  +\Sigma_{\theta}=0, \label{19}%
\end{equation}
where\footnote{It is easy to see that the Liouville equation, $\partial
\hat{\rho}/\partial t+\left(  i/\hbar\right)  [\hat{\rho},\widehat{H}]=0$ when
written in the space coordinate representation acquires the form of Eq.
(\ref{19}).}%
\begin{equation}
\Sigma_{\theta}=-\frac{2R}{\hbar}\operatorname{Im}\left[  e^{-iS/\hbar}%
V\star\left(  R\,e^{iS/\hbar}\right)  \right]  . \label{20}%
\end{equation}
By integrating Eq.\ (\ref{19}) over the space we find
\begin{equation}
\frac{d}{dt}\int R^{2}d^{3}x=0, \label{20.01}%
\end{equation}
since
\begin{equation}
\int\Sigma_{\theta}d^{3}x=0, \label{20.02}%
\end{equation}
and $R^{2}$ vanishes at infinity.

\subsection{Constructing the ontological theory of motion}

The formalism to be\ presented from now on is along the same line as the one
adopted by Bohm and followers (see for example \cite{25,21}). Before
developing the BNCQM, we briefly summarize the essential ideas that lie behind
the Bohmian interpretation.

The Bohmian approach to quantum theory is founded on the assumption that the
complete characterization of a quantum system cannot be provided by a wave
function alone. For the description of individual processes, which are not
statistical in character, an objective view of matter is adopted. In order to
reconcile the\ notion of objective reality with the known results from quantum
theory, an individual physical system is assumed to be\ composed of\ a wave
propagating with a particle. The particle moves under the guidance of the
wave, which satisfies the ordinary Schr\"{o}dinger equation and\ contains the
information on how the energy of the particle must be directed.

As in the commutative counterpart, in the formulation of BNCQM we shall assume
the system as composed of a wave function and a point particle. Contrary to
the commutative case, however, in BNCQM the position observables satisfy the
relation\ (\ref{1}), and the wave function satisfies the noncommutative
Schr\"{o}dinger equation (\ref{14}). Having the equation for the evolution of
the guiding wave $\Psi$, one still has to determine the particle motion. In
principle, there is an arbitrariness in this procedure. However, necessary
conditions for the theory to be capable of reproducing the same statistical
results as the standard interpretation of NCQM constrain\ the admissible form
for the functions $X^{i}(t)$ that describe the particle motion. Notice that
the wave function is valued on canonical coordinates. Therefore, the use of
these coordinates in intermediary calculations to determine the $X^{i}(t)$'s
is unavoidable. Before determining the procedure to find these functions, we
must define the rules for the computation of expectation values in BNCQM.

With an\ arbitrary physical observable, characterized by a Hermitian operator
$\widehat{A}(\hat{x}^{i},\hat{p}^{i})$, it is possible to associate a function
$\mathcal{A}(x^{i},t)$,\ the ``local expectation value'' of $\widehat{A}$
\cite{22},\ which when averaged over the ensemble of density $\rho
(x^{i},t)=\left|  \Psi(x^{i},t)\right|  ^{2}$ gives the same expectation value
obtained by the standard operatorial formalism. It is natural to define the
ensemble average by
\begin{equation}
\langle\widehat{A}\rangle_{t}=\int\rho(x^{i},t)\mathcal{A}(x^{i}%
,t)d^{3}x.\label{20.1}%
\end{equation}
For Eq. (\ref{20.1}) to agree with Eq. (\ref{14.7}), $\mathcal{A}(x^{i},t)$
must be defined as\footnote{Our notation differs from that of Holland
\cite{22} who denotes local expectation value of $\widehat{A}$ by
$A(x^{i},t).$}
\begin{equation}
\mathcal{A}(x^{i},t)=\frac{\operatorname{Re}\left[  \Psi^{\ast}\left(
x^{i},t\right)  \widehat{A}(x^{i},-i\hbar\partial_{i})\Psi\left(
x^{i},t\right)  \right]  }{\Psi^{\ast}\left(  x^{i},t\right)  \Psi\left(
x^{i},t\right)  }=A(x^{i},t)+\mathcal{Q}_{A}(x^{i},t),\label{20.3}%
\end{equation}
where the real value was taken to account for the hermiticity of $\widehat
{A}(\hat{x}^{i},\hat{p}^{i})$ and $A(x^{i},t)=$ $A[x^{i},p^{i}=\partial
^{i}S(x^{i},t)]$, that is, a function obtained from $\widehat{A}(\hat{x}%
^{i},\hat{p}^{i})$ by replacing $\hat{x}^{i}\rightarrow x^{i},\hat{p}%
^{i}\rightarrow\partial^{i}S(x^{i},t)$. $\mathcal{Q}_{A}$ is defined by
\begin{equation}
\mathcal{Q}_{A}=\operatorname{Re}\left[  \frac{\widehat{A}(x^{i}%
,-i\hbar\partial_{i})\Psi\left(  x^{i},t\right)  }{\Psi\left(  x^{i},t\right)
}\right]  -A(x^{i},t)\label{20.4}%
\end{equation}
and is the quantum potential that accompanies $A(x^{i},t)$ (for details of the
procedure to identify quantum effects, see, e.g., \cite{9}). From Eq.
(\ref{20.3}) we find that the local expectation value of Eq. (\ref{14.5}) is
\begin{equation}
X^{i}=x^{i}-\frac{\theta^{ij}}{2\hbar}\partial_{j}S(x^{i},t).\label{22.3}%
\end{equation}
The strategy to find the $X^{i}(t)$'s now\ becomes clear. The relevant
information for particle motion can be extracted from the guiding wave
$\Psi\left(  x^{i},t\right)  $ by first\ computing the associated canonical
position tracks $x^{i}(t)$, and then evaluating Eq. (\ref{22.3}) at
$x^{i}=x^{i}(t)$. In order to find a good equation for the $x^{i}(t)$'s, it is
interesting to consider the Heisenberg formulation and the equations of motion
for the observables (see, for example, \cite{8.96}). For the variables
$\hat{x}^{i}$ they\ are\ given by
\begin{equation}
\frac{d\hat{x}_{H}^{i}}{dt}=\frac{1}{i\hbar}[\hat{x}_{H}^{i},\widehat
{H}]=\frac{\hat{p}_{H}^{i}}{m}+\frac{\theta^{ij}}{2\hbar}\frac{\partial
\widehat{V}(\widehat{X}_{H}^{i})}{\partial\widehat{X}_{H}^{j}}.\label{21}%
\end{equation}

By passing the right-hand side (RHS) of Eq. (\ref{21}) to the Schr\"{o}dinger
picture it is possible to define the velocity operators\
\begin{equation}
\hat{v}^{i}=\frac{1}{i\hbar}[\hat{x}^{i},\widehat{H}]=\frac{\hat{p}^{i}}%
{m}+\frac{\theta^{ij}}{2\hbar}\frac{\partial\widehat{V}(\widehat{X})}%
{\partial\widehat{X}^{j}}. \label{22.2}%
\end{equation}

The differential equation for the canonical positions $x^{i}(t)$ is found by
identifying $dx^{i}(t)/dt$ with the local expectation value\footnote{The
relevance of this procedure for the determination of the equation of motion
will be clearer in the next subsection.} of $\hat{v}^{i}:$%
\begin{equation}
\frac{dx^{i}(t)}{dt}=\left.  \left[  \frac{\partial^{i}S(x^{i},t)}{m}%
+\frac{\theta^{ij}}{2\hbar}\frac{\partial V(X^{i})}{\partial X^{j}}%
+\frac{\mathcal{Q}^{i}}{2}\right]  \right|  _{x^{i}=x^{i}(t)}, \label{22.5}%
\end{equation}
where $X^{i}$ is given in Eq. (\ref{22.3}), $S(x^{i},t)$ is the phase of
$\Psi$, and
\begin{equation}
\mathcal{Q}^{i}=\operatorname{Re}\left(  \frac{\left(  \theta^{ij}%
/\hbar\right)  \left[  \partial\widehat{V}(\widehat{X}^{i})/\partial
\widehat{X}^{j}\right]  \Psi(x^{i},t)}{\Psi(x^{i},t)}\right)  -\frac
{\theta^{ij}}{\hbar}\frac{\partial V(X^{i})}{\partial X^{j}}. \label{23.5}%
\end{equation}
The potentials $\mathcal{Q}^{i}$\ account for quantum effects coming from
derivatives of order 2 and higher contained in $\partial\widehat{V}%
(\widehat{X}^{i})/\partial\widehat{X}^{j}$.

Once the $x^{i}(t)$ are known, the\ particle trajectories are given by
\begin{equation}
X^{i}(t)=x^{i}(t)-\frac{\theta^{ij}}{2\hbar}\partial_{j}S(x^{i}(t),t).
\label{24}%
\end{equation}

One important property of Eq. (\ref{24}) is that the particles' positions are
not defined on nodal regions of $\Psi$, where $S$ is undefined. Thus, the
particles cannot run through these regions. An interesting consequence of this
property is that, although the wave function is valued on the canonical
position variables, its vanishing can be adopted as a boundary condition,
implying that the particles do not run through a region. This is a nontrivial
conclusion, since, as stressed before, $\left.  \left|  \Psi(x^{i},t)\right|
^{2}d^{3}x\right.  $ refers to the canonical variables, and thus\ does not
represent the probability that the particles are in the volume $d^{3}x$ around
the point $\vec{x}$ at time $t$.\ Indeed, it must exclusively be attributed to
the fact that the particles, in the theory under consideration, are objective
and their trajectories are given by Eq.\ (\ref{24}). Had one considered, for
example, the problem of how to apply boundary conditions in NCQM to calculate
the energy levels of a particle in an infinite square well potential from\ the
point of view of the\ orthodox Copenhagen interpretation, there would be no
preferred answer.

The difficulty of introducing well-defined lines\ with boundary conditions in
noncommutative theories was previously stressed in \cite{27.3}. In that
context, noncommutativity was considered as an intrinsic property of the
spacetime. Part of the difficulty in conceiving boundary conditions on
well-defined lines\ is automatically removed if the interpretation for the
noncommutativity proposed in \cite{9} is adopted, since the spacetime in that
work is assumed to be pointwise.\ For the determination of the appropriate
boundary condition for the\ particles not to run through a region in NCQM, the
Bohmian approach is hereby providing the unambiguous prescription one would\ request.

We close this subsection by commenting how the uncertainty (\ref{14.6}) can be
understood in the Bohmian interpretation. In the ordinary de Broglie-Bohm
theory, the impossibility of simultaneously determining the position and
momentum of a particle is attributed to the perturbation introduced on
$p^{i}=\partial^{i}S$ by the evolution of the wave function during the
measurement process \cite{22}. The uncertainty (\ref{14.6}) is generated by a
similar mechanism, since the $X^{i}$'s contain $\partial^{i}S$ in their
definition. Notice that, contrary to the ordinary de Broglie-Bohm theory,
where the initial particle positions can be perfectly known in measurement (by
paying the price of disturbing the system and modifying the wave function),
the initial positions of the particles in BNCQM are experimentally\ undeterminable.

\subsection{The basic postulates}

In the previous subsection, we proposed an objective quantum theory of motion
for NCQM. Let us now\ summarize the complete theory in a formal structure.
This is done with the help of the following postulates.

(1) The spacetime is commutative and has a pointwise manifold structure with
canonical coordinates $x^{i}$. The observables corresponding to operators of
position coordinates ${\hat{X}}^{i}$ of particles satisfy the commutation
relation
\begin{equation}
\lbrack\widehat{X}^{i},\widehat{X}^{j}]=i\theta^{ij}. \label{28}%
\end{equation}
The position observables can be represented in the coordinate space as
$\widehat{X}^{i}=x^{i}+i\theta^{ij}\partial_{j}/2$, and the $x^{i}$ are
canonical coordinates associated with the particle.

(2) A quantum system is composed of a point particle and a wave $\Psi$. The
particle moves in spacetime under the guidance of the wave, which satisfies
the Schr\"{o}dinger equation
\begin{equation}
i\hbar\frac{\partial\Psi(x^{i},t)}{\partial t}=-\frac{\hbar^{2}}{2m}\nabla
^{2}\Psi(x^{i},t)+V(\widehat{X}^{i})\Psi(x^{i},t). \label{29}%
\end{equation}

(3) The particle moves along the trajectory
\begin{equation}
X^{i}(t)=x^{i}(t)-\frac{\theta^{ij}}{2\hbar}\partial_{j}S(x^{i}(t),t)
\label{29.5}%
\end{equation}
independent of observation, where $S$ is the phase of $\Psi$ and\ the
$x^{i}(t)$ describe the canonical position trajectories, which are found by solving%

\begin{equation}
\frac{dx^{i}(t)}{dt}=\left.  \left[  \frac{\partial^{i}S(\vec{x},t)}{m}%
+\frac{\theta^{ij}}{2\hbar}\frac{\partial V(X^{i})}{\partial X^{j}}%
+\frac{\mathcal{Q}^{i}}{2}\right]  \right|  _{x^{i}=x^{i}(t)}\text{.}
\label{30}%
\end{equation}
To find the path followed by a particle, one must specify its initial
canonical position $x^{i}\left(  0\right)  $, solve Eq. (\ref{30}), and then
obtain the physical path via Eq. (\ref{29.5}).

The three postulates presented above constitute on their own a consistent
theory of motion. However, the theory presented is intended to be a finer view
of quantum mechanics, able to give a detailed description of the individual
physical processes and provide the same statistical predictions. In ordinary
commutative Bohmian mechanics, in order to reproduce the statistical
predictions of the Copenhagen interpretation, the additional requirement that,
at a certain instant of time $t_{0}$, $\rho(x^{i},t_{0})=\left|  \Psi(\vec
{x},t_{0})\right|  ^{2}$ is imposed. This assumption and the
\textit{equivariance} \cite{27} of the probability distribution $\rho$ assure
that $\rho(x^{i},t)=\left|  \Psi(x^{i},t)\right|  ^{2}$ for all $t.$ A
distribution $\rho(x^{i},t)=\left|  \Psi(x^{i},t)\right|  ^{2}$ is said to be
\textit{equivariant} if it retains its form as a functional of $\Psi(x^{i},t)$
under evolution of the ensemble particles satisfying ${\dot{x}}^{i}%
(t)=f^{i}(x^{j},t)$. In other words, \textit{equivariance} is achieved if,
departing from an ensemble of physical systems, each one containing a single
particle, whose associated canonical probability density at initial time
$t_{0}$ is\ given by $\rho(x^{i},t_{0})=\left|  \Psi(x^{i},t_{0})\right|
^{2},$ and evolving according to ${\dot{x}}^{i}(t)=f^{i}(x^{j},t)$,
then$\ \rho(x^{i},t)=\left|  \Psi(x^{i},t)\right|  ^{2}$ for all $t.$ In such
a case the probability distribution $\rho(x^{i},t)$ satisfies the transport
equation
\begin{equation}
\frac{\partial\rho}{\partial t}+\frac{\partial(\rho\dot{x}^{i})}{\partial
x^{i}}=0. \label{32}%
\end{equation}

In ordinary commutative\ quantum mechanics\ the equivariance property is
satisfied thanks to the equality $\dot{x}^{i}(t)=J^{i}/\rho$ \cite{27}, which
is a consequence of the identification between $\dot{x}^{i}(t)$ and\ the local
expectation value of the $\hat{v}^{i}$. In the BNCQM proposed in this work,
the same identification is valid. However, this is not sufficient to guarantee
equivariance in all cases. This is rendered evident by computing the canonical
probability current $J^{i}(x^{i},t)$, which is defined by \cite{28}
\begin{equation}
J^{i}(x^{i},t)=\operatorname{Re}\left[  \Psi^{\ast}(x^{i},t)\hat{v}\Psi
(x^{i},t)\right]  =\left|  \Psi(x^{i},t)\right|  ^{2}\left[  \frac
{\partial^{i}S(\vec{x},t)}{m}+\frac{\theta^{ij}}{2\hbar}\frac{\partial
V(X^{i})}{\partial X^{j}}+\frac{\mathcal{Q}^{i}}{2}\right]  =\rho\dot{x}%
^{i},\label{30.5}%
\end{equation}
and regrouping the terms in Eq. (\ref{19})\ in such a way that the canonical
probability flux (\ref{30.5}) appears explicitly,\footnote{The notion of
conserved current in noncommutative theories is little different from the one
in commutative theories, as was pointed out in \cite{25.7}, in the context of
field theory. From the global $U(1)$ symmetry of the noncommutative
Schr\"{o}dinger Lagrangian, the maximum that can be said is that $\partial
\rho/\partial t+\partial J^{i}(x^{i},t)/\partial x^{i}+F_{\theta}=0$, where
$F_{\theta}$ is some function containing the Moyal product and that\ satisfies
$\int d^{3}xF_{\theta}=0.$} obtaining
\begin{equation}
\frac{\partial\rho}{\partial t}+\frac{\partial(\rho\dot{x}^{i})}{\partial
x^{i}}-\frac{\partial}{\partial x^{i}}\left[  \rho\left(  \frac{\theta^{ij}%
}{2\hbar}\frac{\partial V(X^{i})}{\partial X^{j}}+\frac{\mathcal{Q}^{i}}%
{2}\right)  \right]  +\Sigma_{\theta}=0.\label{31.5}%
\end{equation}
For equivariance to occur, an additional condition that the sum of the last
two terms in the RHS of Eq. (\ref{31.5}) vanishes is required. When $V(X^{i})$
is a linear or quadratic function, as in the application problem of the next
section, such a condition is trivially satisfied,\footnote{This is easily seen
by substituting\ Eq.\ (\ref{14.5}) in\ Eq.\ (\ref{29}), regrouping the terms,
and noticing that in these cases $\widehat{H}$ is reduced to a ``familiar''
Hamiltonian quadratic in the canonical momenta, as occurs in commutative
quantum mechanics. The noncommutative effects, however, are still present, as
we shall show in the next section.} and thus $\rho(x^{i},t)=\left|  \Psi
(x^{i},t)\right|  ^{2}$ is certainly equivariant.\ The same\ may also occur
for special states when other potentials are considered in (\ref{29}), but it
is not a general property of (\ref{31.5}). We shall return to this point in
the final discussion.

\section{Bohmian Noncommutative Harmonic Oscillator}

Here, we show a simple application of the BNCQM for the analysis of a
two-dimensional harmonic oscillator. We shall follow the approach previously
discussed in \cite{9}. Other relevant work on the noncommutative harmonic
oscillator may be found in \cite{8.95}. In two dimensions, (\ref{1}) is
reduced to
\begin{equation}
\lbrack\widehat{X}^{\mu},\widehat{X}^{\nu}]=i\theta\epsilon^{\mu\nu}.
\label{34}%
\end{equation}
The position observables of the particles can therefore be represented by
$\widehat{X}^{i}=x^{i}-\theta\epsilon^{ij}\hat{p}_{j}/2\hbar$, and the
harmonic oscillator Hamiltonian is written as
\begin{equation}
H=\frac{1}{2m}\left(  \hat{p}_{x}^{2}+\hat{p}_{y}^{2}\right)  +\frac{1}%
{2}mw^{2}\left[  \left(  x-\frac{\theta}{2\hbar}\hat{p}_{y}\right)
^{2}+\left(  y+\frac{\theta}{2\hbar}\hat{p}_{x}\right)  ^{2}\right]  ,
\label{35}%
\end{equation}
where $m$ and $w$ are the mass and frequency of the associated commutative
oscillator, respectively.

The corresponding Schr\"{o}dinger equation in polar coordinates is
\begin{align}
i\hbar\frac{\partial\Psi_{\theta}\left(  r,\varphi,t\right)  }{\partial t}  &
=H_{\theta}\Psi_{\theta}\left(  r,\varphi,t\right) \nonumber\\
&  =-\frac{\hbar^{2}}{2m}\left[  1+\left(  \frac{mw\theta}{2\hbar}\right)
^{2}\right]  \left(  \partial_{r}^{2}+\frac{1}{r}\partial_{r}+\frac{1}{r^{2}%
}\partial_{\varphi}^{2}\right)  \Psi_{\theta}\left(  r,\varphi,t\right)
\label{36}\\
&  +\left(  i\frac{m}{2}\theta w^{2}\partial_{\varphi}+\frac{m}{2}w^{2}%
r^{2}\right)  \Psi_{\theta}\left(  r,\varphi,t\right)  ,\nonumber
\end{align}
whose solution is \cite{9}
\begin{equation}
\Psi_{\theta}\left(  r,\varphi,t\right)  =\left(  -1\right)  ^{n}\sqrt
{\frac{n!\tilde{\zeta}}{\pi\left(  n+\left|  \alpha\right|  \right)  !}}%
\exp\left(  -\frac{\tilde{\zeta}r^{2}}{2}\right)  \left(  \sqrt{\tilde{\zeta}%
}r\right)  ^{\left|  \alpha\right|  }L_{n,\theta}^{^{\left|  \alpha\right|  }%
}\left(  \tilde{\zeta}r^{2}\right)  e^{i\alpha\varphi-iEt/\hbar}, \label{36.5}%
\end{equation}
where $L_{n,\theta}^{^{\left|  \alpha\right|  }}\left(  \tilde{\zeta}%
r^{2}\right)  $ are the associated Laguerre polynomials
\begin{equation}
L_{n,\theta}^{^{\left|  \alpha\right|  }}\left(  \tilde{\zeta}r^{2}\right)
=\overset{n}{\underset{l=0}{\sum}}\left(  -1\right)  ^{l}\left(
\begin{array}
[c]{c}%
n+\left|  \alpha\right| \\
n-l
\end{array}
\right)  \frac{\left(  \tilde{\zeta}r^{2}\right)  ^{l}}{l!},\text{
\ \ \ }\tilde{\zeta}^{2}=\frac{\left(  mw/\hbar\right)  ^{2}}{1+\left(
mw\theta/2\hbar\right)  ^{2}}, \label{36.7}%
\end{equation}
$n=0,1,2,$... is the principal quantum number, and $\alpha=0,\pm1,\pm2,$... is
the canonical angular momentum quantum number.

The energy levels are given by
\begin{equation}
E_{n,\alpha,\theta}=2\hbar w\left[  1+\left(  \frac{mw\theta}{2\hbar}\right)
^{2}\right]  ^{1/2}\left(  n+\frac{\left|  \alpha\right|  +1}{2}\right)
-\frac{m\theta w^{2}\alpha}{2}. \label{36.9}%
\end{equation}
Notice that, due to the noncommutative effects, the degeneracy of the energy
levels corresponding to\ the right- and left-handed polarizations for the same
$n$ is removed. When the noncommutativity (\ref{1}) is assumed as originating
from the action of a strong background field, like the Neveu-Schwartz field in
the stringy context \cite{1}, or a magnetic field when a condensed matter
system is projected onto its lowest Landau level, the lifting of the
degeneracy can be \ intuitively understood as the consequence of a chirality
introduced by the background field.

For simplicity, let us consider the state where $n=1$. In\ this state, Eq.
(\ref{36.5}) is simplified to
\begin{equation}
\Psi_{\theta}\left(  r,\varphi,t\right)  =\sqrt{\frac{\tilde{\zeta}}%
{\pi\left|  \alpha\right|  !}}\exp\left(  -\frac{\tilde{\zeta}r^{2}}%
{2}\right)  \left(  \sqrt{\tilde{\zeta}}r\right)  ^{\left|  \alpha\right|
}\left(  1+\left|  \alpha\right|  -\tilde{\zeta}r^{2}\right)  e^{i\alpha
\varphi-iEt/\hbar}, \label{37}%
\end{equation}
and the corresponding $V,V_{nc},$ $Q_{c},$ $Q_{nc},$ and $\Sigma_{\theta}$
are
\begin{align}
V  &  =\frac{1}{2}mw^{2}r^{2},\nonumber\\
V_{nc}  &  =\left(  \frac{mw\theta}{2\hbar}\right)  ^{2}\frac{\alpha^{2}%
\hbar^{2}}{2mr^{2}}-\frac{m\theta w^{2}\alpha}{2},\nonumber\\
Q_{c}  &  =-\frac{1}{2}mw^{2}r^{2}+\hbar w\left(  \left|  \alpha\right|
+3\right)  -\frac{\alpha^{2}\hbar^{2}}{2mr^{2}},\label{38}\\
Q_{nc}  &  =\left[  \sqrt{1+\left(  \frac{mw\theta}{2\hbar}\right)  ^{2}%
}-1\right]  \hbar w\left(  \left|  \alpha\right|  +3\right)  -\left(
\frac{mw\theta}{2\hbar}\right)  ^{2}\frac{\alpha^{2}\hbar^{2}}{2mr^{2}%
},\nonumber\\
\mathcal{Q}^{i}  &  =\Sigma_{\theta}=0.\nonumber
\end{align}

The canonical trajectories are found by solving the equations
\begin{equation}
\frac{dx}{dt}=\frac{1}{m}\frac{\partial S}{\partial x}+\frac{\theta}{2\hbar
}mw^{2}\left(  y+\frac{\theta}{2\hbar}\frac{\partial S}{\partial x}\right)
\text{, \ \ \ }\frac{dy}{dt}=\frac{1}{m}\frac{\partial S}{\partial y}%
-\frac{\theta}{2\hbar}mw^{2}\left(  x-\frac{\theta}{2\hbar}\frac{\partial
S}{\partial y}\right)  .\label{39}%
\end{equation}
Changing into polar coordinates and substituting $S=\alpha\hbar\varphi-Et$ in
Eq. (\ref{39}), we find
\begin{equation}
\frac{dr}{dt}=0\text{,\ \ \ \ }\frac{d\varphi}{dt}=\frac{\hbar\alpha}{mr^{2}%
}+\frac{mw^{2}\theta^{2}\alpha}{4\hbar r^{2}}-\frac{mw^{2}\theta}{2\hbar
},\label{40}%
\end{equation}
whose solutions are
\begin{equation}
r=r_{0}\text{, \ \ \ \ \ }\varphi=\varphi_{0}+w_{\theta}t\text{,
\ \ \ }w_{\theta}=\left(  \frac{\hbar\alpha}{mr^{2}}+\frac{mw^{2}\theta
^{2}\alpha}{4\hbar r^{2}}-\frac{mw^{2}\theta}{2\hbar}\right)  .\label{41}%
\end{equation}
The physical radius and angle are
\begin{align}
R(t) &  =\sqrt{X^{2}(t)+Y^{2}(t)}=\left[  \left(  x-\frac{\theta}{2\hbar}%
\frac{\partial S}{\partial y}\right)  ^{2}+\left(  y+\frac{\theta}{2\hbar
}\frac{\partial S}{\partial x}\right)  ^{2}\right]  ^{1/2}\nonumber\\
&  =r_{0}\left|  1-\frac{\alpha\theta}{2r_{0}^{2}}\right|  =R_{0}\label{42}%
\end{align}
and
\begin{equation}
\Phi(t)=\arctan\left[  \frac{Y(t)}{X(t)}\right]  =\arctan\left[  \frac
{y(t)}{x(t)}\right]  =\varphi\left(  t\right)  .\label{43}%
\end{equation}
The velocity of the particles is tangential to their circular orbits, being
given by
\begin{equation}
\left|  \mathbf{V}\left(  t\right)  \right|  =\sqrt{\dot{X}^{2}(t)+\dot{Y}%
^{2}(t)}=R_{0}\left|  \dot{\varphi}\right|  =R_{0}\left|  w_{\theta}\right|
.\label{44}%
\end{equation}
As expected, the particle trajectories are circles. Contrary to the
commutative case, however, the absolute value of the tangential velocity
$\left|  \mathbf{V}\left(  t\right)  \right|  $ is not the same for both the
right- and left-polarized states. This is due to a difference in the angular
velocity $w_{\theta}$ and in the radius $R_{0}$ of their corresponding orbits.
While the orbits associated with the right-handed excitations have the energy
levels shifted downward, and their velocities and radii reduced with respect
to the commutative ones, the left-handed excitations have their energy levels
shifted upward, and move with larger velocities and radii. Notice from
Eq.\ (\ref{44}) that, when the system is in the lowest-energy state,
characterized by $\alpha=0$, the particle is still moving,\footnote{This is
not the source of any inconsistency. Since $\alpha$ is\ a quantum number
related to the canonical position variables, it is not directly connected to
the physical angular momentum.} unless $R_{0}=0.$ Such a motion is absent in
ordinary Bohmian theory and originates from a noncommutative term contained in
$w_{\theta}$.

From Eq. (\ref{38}), it is possible to see that the condition $V_{nc}%
+Q_{nc}\rightarrow0$\ is satisfied if $\theta\ll2\hbar/mw$, as shown in
\cite{9}. In that work, there was also an\ assumption that $\theta$ should be
sufficiently small in order that Eq. (\ref{14.6}) could not be directly
verified by experimentation. The length scales considered until now,
therefore, were assumed to be many times larger than that of $\sqrt{\theta},$
which is the characteristic length of noncommutativity. On scales of
$\sqrt{\theta}$ order or smaller, noncommutativity effects associated with Eq.
(\ref{14.6})\ are expected to drastically modify the behavior of the system
\cite{3,6}. Let us ignore, for the time being, the previous assumption on the
dimensions of our system, and allow the noncommutative harmonic oscillator to
live at arbitrarily small length\ scales, or, equivalently, allow $\theta$ to
assume a\ large value. In this case, it is interesting to consider the
individual behavior of $V_{nc}$ and $Q_{nc}$. To study the oscillator orbits,
we compare the behavior of the variable$\ R,$ which describes\ its physical
radius, with that of the canonical variable $r$. A plot of $R(r)$ is found in
Fig. 1 for the cases where $\alpha\theta=1$ and $\alpha\theta=-1$.%

\begin{center}
\includegraphics[
trim=0.035866in 0.035938in 0.000000in 0.036237in,
height=8.2373cm,
width=11.1039cm
]%
{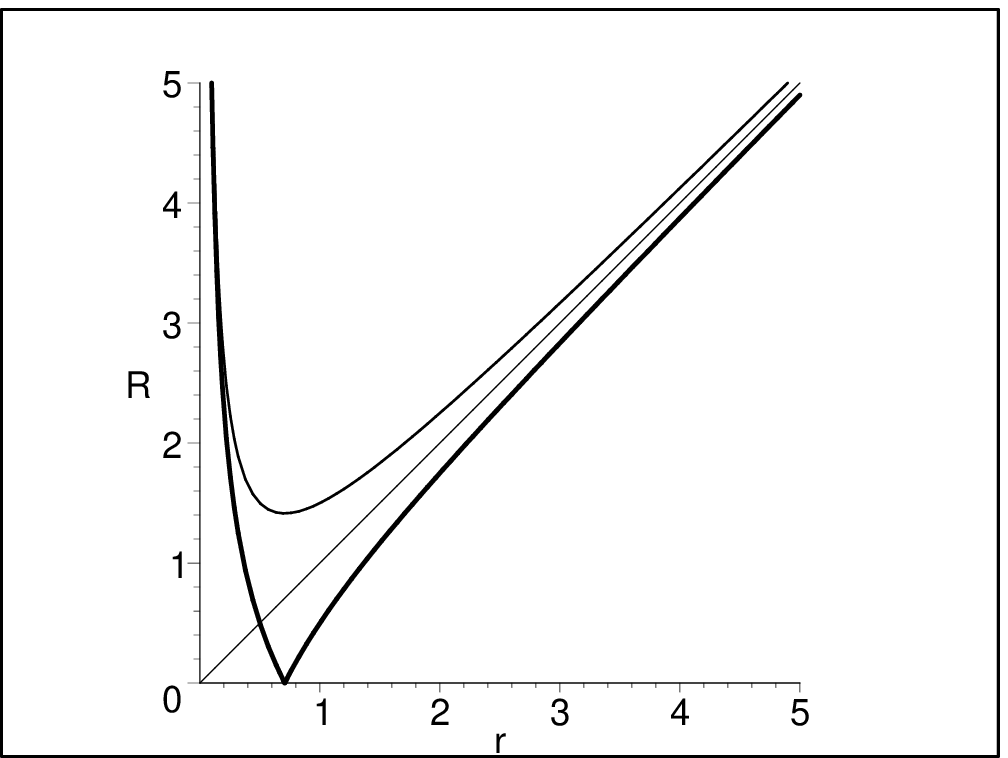}%
\\
\textbf{FIG. 1}. The typical\ behavior of the radius of the oscillator orbit,
$R(r),$ for $\alpha>0$ (thick line) and $\alpha<0$ (thin line) illustrated for
the cases where $\alpha\theta=1$ and $\alpha\theta=-1$.\bigskip
\end{center}
\bigskip

From Fig. 1 we can see that, when\ the scale of the system is sufficiently
large as compared to $\sqrt{\left|  \alpha\theta\right|  }$, the\ distinction
between $r$ and $R$ is not relevant. However, for length scales around
$\sqrt{\left|  \alpha\theta\right|  }$, the distinction between $R$ and $r$
begins to become important, since these variables can differ by a large
amount. The asymmetry between the states for which $\alpha>0$ and $\alpha<0$
is also important at these scales. If $\alpha>0,$ there is a specific orbit
radius in which canonical and physical coordinates are completely identified,
corresponding to the point where the straight line $R=r$ crosses the
$\alpha>0$ thick line. From Eq. (\ref{42}), it is easy to see that this point
is $r_{0}=\sqrt{\alpha\theta}/2$. By substituting this value for $r$ in Eq.
(\ref{38}), one finds $V_{nc}=0$. Since $Q_{nc}\neq0,$\ one still has
noncommutative effects in the corresponding orbit, but they have a\ genuinely
quantum nature. The minimum value allowed for $R_{0}$ occurs when $r_{0}%
=\sqrt{\left|  \alpha\theta\right|  /2},$ for both $\alpha>0$ and $\alpha<0$
states. Its value is$\ R_{0M}=0$\ for $\alpha>0$, and $R_{0M}=\sqrt{2\left|
\alpha\theta\right|  }$ for $\alpha<0$.

There is an interesting property of the noncommutative harmonic oscillator
that is rendered evident by its Bohmian description. Observe in Fig. 1 that,
for each value of$\ R,$ there correspond two values of $r$, one smaller than
$\sqrt{\left|  \alpha\theta\right|  /2}$ and the\ other larger. The smallest
values of $r$\ contained in the interval $(0,$\ $\sqrt{\left|  \alpha
\theta\right|  /2})$ have as their partners exactly the largest ones in the
interval $(\sqrt{\left|  \alpha\theta\right|  /2},\infty)$. In the ordinary
commutative\ Bohmian interpretation, there is a contextual character in the
information provided by the wave function to guide the particle motion (for
details see, for example, \cite{22}). The information for the particle to
move, however, is extracted by the particle from the part of the wave function
that is spread over the spatial points which cover the particle position and a
small neighborhood around it.\ What we find in the noncommutative Bohmian
harmonic oscillator is a new and interesting property, over and above the
contextuality of ordinary Bohmian mechanics. For the case where the system has
a large physical radius, it may be receiving information from part of the wave
function that is concentrated in a small region far beyond the position where
the particle is located. This property is similar to the IR-UV mixing that
occurs in field theory \cite{5} and is a manifestation of the specific kind of
nonlocality found in NCQM due to the ``shift'' in the interaction point in Eq.
(\ref{14}).

\section{Discussion and Outlook}

In this work we carried out an investigation of the possibility of developing
a Bohmian interpretation for NCQM. The theory was proposed to be based on a
commutative spacetime containing point particles, whose position observables
should satisfy the canonical relation (\ref{1}). Such a realization of
noncommutativity only between particle position observables means that,
although the particles are point like, their complete localization in a
process of measurement is forbidden by the disturbances caused by the
measurement apparatus interacting with the quantum system. The intrinsic
uncertainty of the particle localization during a measurement process must be
faced on the same footing as the one forbidding the simultaneous determination
of the momentum and the position of the particle in ordinary Bohmian\ theory.

In the manner it was constructed, BNCQM was conceived\ to reproduce the same
statistical predictions of NCQM in the Copenhagen interpretation with a
continuous evolution law for particle motion. As a result of\ our
investigation of the possibility for this to occur, we found that, when linear
and quadratic potentials are considered in the Hamiltonian, the theory in its
present form is certainly statistically equivalent to NCQM in the Copenhagen
interpretation. The same may also occur for other specific potential terms or
physical states where Eqs.\ (\ref{32}) and (\ref{31.5}) do match. When this is
not the case, for the theories' predictions to agree, it is necessary to
modify the particle evolution law. One interesting way this may be done is by
considering a ``hybrid'' evolution law constituted of a continuous part
governed by a differential equation added to a stochastic piece, which allows
the particles to jump, along the lines adopted in \cite{40}. Equation
(\ref{31.5}) in this case should be understood as a probability transport
equation of a piecewise-deterministic jump process.

Among the physical systems of interest where BNCQM with a continuous particle
motion is able to reproduce the Copenhagen interpretation results is the
harmonic oscillator. Although very specific, the harmonic potential is of
great relevance for physics, which justifies the enormous variety of work on
NCQM devoted to it (see \cite{8.95} and references therein). Indeed, in this
work, the harmonic oscillator proved to be useful to illustrate essential
properties of NCQM in the spirit of the Bohmian interpretation. In its fine
description of the harmonic oscillator, BNCQM revealed the interesting
possibility that the small-scale physics can influence the large-scale
phenomena in the quantum-mechanical context, in close similarity to the IR-UV
mixing that appears in field theory. Another important contribution of the
Bohmian interpretation comes from its capability to give well-defined
predictions in situations where the Copenhagen interpretation is vague, as
discussed in Sec. III.

An interesting environment where the predictions of the Bohmian interpretation
may be confronted in future with experimentation is that of quantum cosmology.
Recently, noncommutativity at early times of the universe was introduced by
deforming the commutation relation among the minisuperspace variables\ in a
cosmological model based on the Kantowski-Sachs metric \cite{20.7},
originating a noncommutative Wheeler-DeWitt equation. Since in the formalism
of minisuperspace the Wheeler-DeWitt equation is essentially quantum
mechanical, the application of the Bohmian interpretation developed in this
work for models like the one of \cite{20.7} is almost immediate \cite{41}. It
may reveal unknown features or unexpected results, since it will allow a
description of the primordial quantum universe following a well-defined
``trajectory'' in minisuperpace. In\ the\ case of conceiving a quantum
cosmology based on the canonical noncommutativity of the spatial coordinates
(\ref{1}), for example, the ideas presented in this work may also be a good
starting point.

Since the ontological interpretations have variants and are still under
construction, this work should not be considered a closed structure. Many of
the rules stated\ here are open and may be subject to reformulation after
further discussion. In addition to the interesting possibility for a Bohmian
description with stochastic particle jumps to shed light on the interpretation
of all terms in Eq. (\ref{31.5}), which compels us to carry on an extension of
the theory presented here, there are many open questions to be exploited in
the formulation. Among them, we quote the extension of the theory to
incorporate a many-body approach, where some care must be taken when
considering charged particles \cite{8.8}, for example.

\section*{Acknowledgments}

The authors are greatly indebted to Jos\'{e} Hela\"{y}el-Neto and Roderich
Tumulka\ for relevant comments and for all the\ corrections\ to earlier
versions of the manuscript.\ They also acknowledge Jos\'{e} Acacio de
Barros\ for useful discussions highlighting important aspects of the Bohmian
interpretation. This work was financially supported by CAPES and CNPq.


\begin{thebibliography}{99}
\bibitem{1}N. Seiberg and E. Witten, J. High Energy Phys.\textit{
}\textbf{09,} 032\ (1999).

\bibitem{2}R. J. Szabo, Phys. Rep. \textbf{378, }207 (2003).

\bibitem{3}M. R. Douglas and N.A.~Nekrasov, Rev. Mod. Phys. \textbf{73,} 977 (2002).

\bibitem{5}S.~Minwalla, M.~Van~Raamsdonk and N.~Seiberg, J. High Energy
Phys.\textit{ }\textbf{02,} 020 (2000).

\bibitem{6}I.~Bars, ``Nonperturbative Effects of Extreme Localization in
Noncommutative Geometry'', hep-th/0109132.

\bibitem{7}S. M. Carroll, J. A. Harvey, V.A.~Kostelecky, C.D. Lane and T.
Okamoto, Phys. Rev. Lett. \textbf{87,} 141601 (2001).

\bibitem{8}G.~Amelino-Camelia, L.~Doplicher, S. Nam and Y.-S. Seo, Phys. Rev.
\textbf{D 67, }085008 (2003);

I.~Hinchliffe and N.~Kersting, Int. J. Mod. Phys. \textbf{A 19,} 179\ (2004);

G.~Amelino-Camelia, G.~Mandanici and K.~Yoshida, J. High Energy Phys.\textit{
}\textbf{01,} 037 (2004);

A. Anisimov, T. Banks, M. Dine, M. Graesser,\ Phys. Rev. \textbf{D 65} ,
085032 (2002).

\bibitem{8.5}M.~Chaichian, M. M. Sheikh-Jabbari and A.~Tureanu, Phys. Rev.
Lett. \textbf{86, }2716 (2001).

\bibitem{8.7}M. Chaichian, M. M. Sheikh-Jabbari, A. Tureanu\textit{, }hep-th/0212259;

H. Falomir, J. Gamboa, M. Loewe, F. M\'{e}ndez and J. C. Rojas\textit{, }Phys.
Rev.\textit{ }\textbf{D 66, }045018 (2002);

M. Haghighat, S. M. Zebarjad, F. Loran\textit{, }Phys. Rev. \textbf{D 66,
}016005 (2002);

M. Caravati, A. Devoto, W. W. Repko\textit{,} Phys. Lett. \textbf{B 556, }123 (2003);

M. Haghighat, F. Loran\textit{, }Phys. Rev. \textbf{D 67, }096003 (2003)\textit{.}

\bibitem{8.8}P-M. Ho and H-C. Kao \textit{,} Phys. Rev. Lett. \textbf{88
}151602 (2002).

\bibitem{8.9}V. P. Nair and A. P. Polychronakos,Phys. Lett. \textbf{B 505,
}267 (2001);

B. Morariu and A. P. Polychronakos, Nucl. Phys. \textbf{B 610}, 531 (2001);

H. R. Christiansen, F. A. Schaposnik, Phys. Rev. \textbf{D 65}, 086005 (2002).

\bibitem{8.95}M. Demetrian and D. Kochan, Acta Physica Slovaca \textbf{52}, 1 (2002);

R. Banerjee, Mod. Phys. Lett. \textbf{A 17}, 631 (2002);

J. Gamboa, M. Loewe, J. C. Rojas, Phys. Rev. \textbf{D 64,} 067901 (2001);

V.P. Nair and A.P. Polychronakos, Phys. Lett. \textbf{B 505}, 267 (2001);

B. Morariu and A.P. Polychronakos, Nucl. Phys. \textbf{B 610}, 531 (2001);

B. Muthukumar, P. Mitra, Phys. Rev. \textbf{D 66}, 027701 (2002);

S. Bellucci, A. Nersessian, Phys. Lett. \textbf{B 542,} 295 (2002);

C. Acatrinei, J. Phys. \textbf{A 37,} 1225\ (2004).

\bibitem{8.96}C. Acatrinei, J. High Energy Phys\textit{. }\textbf{09, }007 (2001).

\bibitem{9}G. D. Barbosa\textit{,} J. High Energy Phys. \textbf{05}, 024 (2003).

\bibitem{9.3}D. Bohm, Phys. Rev. \textbf{85}, 166 (1952); Phys. Rev.
\textbf{85}, 180 (1952).

\bibitem{9.4}G. 't Hooft, ``Determinism and Dissipation in Quantum Gravity'',
Erice 1999,\textit{ Basics and Highlights in Fundamental Physics}, (1999)\ p.
397, hep-th/0003005;

G. 't Hooft, Class. Quant. Grav. \textbf{16}, 3263 (1999).

\bibitem{17}F. Lalo\"{e}\textit{,} Am. J. Phys\textit{.}\textbf{\ 69,} 655 (2001).

\bibitem{16.5}M. A. B. Whitaker\textit{, }Progr. Quantum Electron.\textit{
}\textbf{24, }1\textbf{ }(2000).

\bibitem{17.5}D. A. Eliezer and R. P. Woodard, Nucl. Phys. \textbf{B 325}, 389 (1989).

\bibitem{9.5}L. Smolin, Matrix Models as Non-Local Hidden Variables Theories ,
\textit{Fukuoka 2001, String theory}, (2001), p. 244, hep-th/020103.

\bibitem{11}D. D\"{u}rr, S. Goldstein, R. Tumulka, N. Zangh\`{i}, J. Phys.
\textbf{A 36}, 4143 (2003).

\bibitem{12}D. Home and A. S. Majumdar, Found. Phys.\textbf{\ 29}, 721 (1999).

\bibitem{14}L. Delle Site, Europhys. Lett. \textbf{57}, 20 (2002);

A. S. Sans, F. Borondo and S. Miret-Art\'{e}s,\textit{\ }J. Phys.:Condens.
Matter \textbf{14}, 6109 (2002).

\bibitem{15}R. H. Parmenter and R. W. Valentine, Phys. Lett.\textit{
}\textbf{A 201}, 1 (1995);\textbf{ 227}, 5 (1997).

\bibitem{16}A. Legget\textit{,} J. Phys.\textit{: }Condens. Matter
\textbf{14}, R415 (2002) R415;

T. Calarco, M. Civi and R. Onofrio\textit{,} J. Supercond. \textbf{12}, 819 (1999);

A. J. Leggett and A. Garg\textit{,} Phys. Rev. Lett. \textbf{54}, 857 (1985).

\bibitem{18}C. R. Leavens, in \textit{Bohmian Mechanics and Quantum Theory:an
Appraisal,} edited by J. T. Cushing, A.Fine, and\ S.Goldstein (Kluwer,
Dordrecht, 1996).

\bibitem{19}M. Daumer, D. D\"{u}rr, S. Goldstein, N. Zangh\`{i}\textit{,} J.
Stat. Phys.\textit{ }\textbf{88}, 967(1997) 967.

\bibitem{20}D. D\"{u}rr, S. Goldstein, S. Teufel, N. Zangh\`{i}\textit{,}
Physica \textbf{A 279}, 416 (2000).

\bibitem{20.5}V. Allori and N. Zangh\`{i}, ``What is Bohmian
Mechanics'',\textit{ }quant-ph/0112008.

\bibitem{4}S. Doplicher, K. Fredenhagen and J. E. Roberts, Phys. Lett.
\textbf{B 331} 39(1994); Comm. Math. Phys. \textbf{172, }187 (1995).

\bibitem{22.5}S. de Haro, \textit{Class. }Quant. Grav. \textbf{15}, 519 (1998).

\bibitem{20.7}H. Garcia-Compe\'{a}n,\textit{\ }O. Obreg\'{o}n and C.
Ram\'{i}rez\textit{, }Phys. Rev. Lett. \textbf{88}, 161301 (2002).

\bibitem{23}R. P. Feynman, F. B. Morinigo, and W. G. Wagner, \textit{Feynman
Lectures on Gravitation, (}Addison-Wesley, Reading, MA, 1995).

\bibitem{23.2}N. Pinto Neto, \textit{Quantum Cosmology}, \textit{VIII
Brazilian School of Cosmology and Gravitation}, (Editions Frontieres,
Gif-sur-Yvette, 1996); CBPF-NF-006-97 [http://www.biblioteca.cbpf.br/index\_2.html];

N. Pinto-Neto and E. S. Santini\textit{, }Phys. Rev. \textbf{D 59, }123517 (1999).

\bibitem{23.5}F. G. Alvarenga, A. B. Batista, J. C. Fabris, S. V. B.
Goncalves, ``Anisotropic Quantum Cosmological Models: A Discrepancy Between
Many-Worlds and dBB Interpretations'', gr-qc/0202009;

N. Pinto-Neto, E. S. Santini, Phys. Lett. \textbf{A 315}, 36 (2003);

R. Colistete Jr., J. C. Fabris, and \ N. Pinto-Neto, Phys. Rev. \textbf{D 62},
083507 (2000).

\bibitem{35}D.~Bigatti and L.~Susskind, Phys. Rev.\textit{ }\textbf{D 62,
}066004 (2000).

M.M. Sheikh-Jabbari, Phys. Lett. \textbf{B 455}, 129 (1999).

\bibitem{25}P. R. Holland, Phys. Rep. \textbf{224,} 95 (1993).

\bibitem{21}D. Bohm, B. J. Hiley and P. N. Kaloyerou\textit{, }Phys. Rep.
\textbf{144, }349 (1987).

D. Bohm, B. J. Hiley, \textit{The Undivided Universe: An Ontological
Interpretation of Quantum Theory,} London (Routledge \& Kegan Paul, London,1993).

\bibitem{22}P. R. Holland, \textit{The Quantum Theory of Motion: An Account of
the de Broglie-Bohm Causal Interpretation of Quantum Mechanics (}World
Scientific, Singapore, March 1998).

\bibitem{27.3}M.~Chaichian, A.~Demichev, P.~Pre\v{s}najder, M.M.
Sheikh-Jabbari and A.~Tureanu, Nucl. Phys. \textbf{B 611}, 383 (2001).

\bibitem{27}D. D\"{u}rr, S. Goldstein and N. Zangh\`{i}, J. Stat.
Phys.\textbf{67}, 843 (1992).

\bibitem{25.7}A.~Micu and M.M. Sheikh~Jabbari, J. High Energy Phys.
\textbf{01}, 025 (2001).

\bibitem{28}L. E. Ballentine, \textit{Quantum Mechanics: A Modern Development
(}World Scientific, Singapore, 1998).

\bibitem{40}D. D\"{u}rr, S. Goldstein, R. Tumulka, N. Zangh\`{i}, ``Quantum
Hamiltonians and Stochastic Jumps''\textit{, }quant-ph/0303056.

\bibitem{41}G. D. Barbosa and N. Pinto-Neto, work in progress.
\end{thebibliography}
\end{document}